# Evolution of surfactant-free "pristine" emulsions.

*Andrei Dukhin[1], Renliang Xu[2], Darrell Velegol[3]*

[1] adukhin@dispersion.com, Dispersion Technology Inc, 364 Adams Street, Bedford Hills, NY 10507

[2] renliang.xu@yahoo.com, 13084 NW 13 Street, Pembroke Pines, FL 33028

[3] velegol@psu.edu, Department of Chemical Engineering, Penn State University, *CBEB 225, University Park PA 16802*

**Abstract**

The term "pristine interface" was introduced by Beattie and Djerdjev 20 years ago (*Angew.Chem.Int.Ed.43,2004*) for emulsions that consist of only water and oil with no surfactant. They are different from the Pickering emulsions, which are also surfactant-free but stabilized with colloidal particles. We overview existing literature on such emulsions and list a wide variety of liquids capable of creating such stable oil-in-water and water-in-oil emulsions.

In contrast to previous studies, we monitor the kinetics of the initial stages of emulsion formation. We conducted such tests in "open setup" when samples are open to air and $CO_2$ content in water varies, and in "closed setup" when samples are isolated and $CO_2$ content is fixed.

For the "open setup", we discovered that sonication and initial pH adjustment to pH range above 9 leads to the emulsions with high zeta potential and sub-micron droplet size. There are two evolution patterns: short- and long-terms. The "short term" lasts about one day and associated with declining pH and increasing zeta potential, whereas droplet size remains almost constant. The "long term" is unraveling during several days, even weeks, exhibit only droplet size increase toward saturation value, with pH and zeta potential remaining constant. The rate of this increase is dependent on mixing conditions.

Emulsification at the "closed setup" is much less pronounced and pH remain constant. This difference supports importance of adsorbed $CO_2$ and related carbonic acid anions in the formation of pristine emulsions and charging droplets interfaces.

We hypothesize the existence of a layer of structured water molecules at the interface, following Eastoe and Ellis (*Adv in Colloid and Interface Sci., 134-135, 89-95, 2007*). Then we point out that Electric Double Layer exerts a force on the water dipole moments in this layer (dielectrostatic force) that compensates Kelvin's pressure. The droplet size calculated using this model is close to our measured sizes. The second factor associated with this layer is the repulsion of the water dipole moments, which we show can compensate for surface tension parallel to the interface. After ruling out alternative hypotheses with our data, we conclude that the model suggested for explaining the stability of nano-bubbles is also consistent with our results for these "pristine emulsions".

**Introduction**

There is an unusual type of emulsion described by several independent groups from different countries during the last 25 years [1-18]. These emulsions consist of only two liquids with no added surfactant. They are different from Pickering emulsions [19], which are stabilized with colloid particles. There is a term "surfactant-free emulsions" that is applied for classifying both of them, which might create confusion. This type of emulsion does not have any added surface stabilizing agent, in contrast to the Pickering emulsions. Perhaps this was a motivation for Beattie and Djerdjev to introduce a special term for such a liquid-liquid interface – "pristine" [10]. It would be logical to call such emulsions "pristine emulsions" for distinguishing them from Pickering emulsions [19].

This type of emulsion is very promising apparently for controlling chemical kinetics according to the extensive review by LaCour et all [20]. This review describes examples of many organic and redox reactions that occur with much faster kinetics in water microdroplets and oil−water emulsions than in bulk solution.

We have summarized all published papers that we knew or could find on this subject in Table 1, listing all liquids used for making such emulsions. This list is quite diverse, which allows us to hypothesize that there is a common mechanism leading to these emulsions' stability. One might assume that the lifespan of such "pristine emulsions" would be very short. They would be destroyed either by coalescence, or by Oswald ripening, or both. Surprisingly, they turn out to be much more stable. Authors of the cited papers express consensus in the long-term stability of these emulsions, lasting days and weeks. This paradoxical long-term longevity of pristine emulsions is similar to dispersions of nano-bubbles [21,22].

**Table 1. Summary of publications on the "pristine emulsions". All emulsions consist of water and oil. The oil phase might be different. We list these oils in the third column that is labeled "oil".**

| Year | Country (and affiliation) of authors [ reference] | Oil |
|------|------|------|
| Oil-in-water emulsions | | |
| 1999-2018 | Japan [1-6] | <ul><li>Hexadecane</li><li>Hexane</li><li>Benzene</li><li>Oleic acid</li><li>Esters of oleic acid</li></ul> |
| 2003-2004 | University of California, USA [7,8] | <ul><li>Dodecane</li><li>Hexane</li><li>Octane</li><li>Decane</li><li>Octadecane</li><li>Squalane</li><li>Hexamethylsqualane</li><li>4-fluorotoluene</li></ul> |
| 2004 | Bristol University, UK [9] | <ul><li>Dodecane</li></ul> |
| 2004, 2025 | Beattie and Djerdjev from Sydney University, Australia [10,11] | <ul><li>Hexadecane</li><li>Decane</li><li>Dodecane</li><li>Eicosane</li><li>Squalane</li></ul> |
| 2020 | China [17] | <ul><li>Hexadecane</li></ul> |
| 2022 | Korea [12] | <ul><li>Olive oil</li></ul> |
| 2023 | China, Japan, Canada [16] | <ul><li>Decane</li></ul> |
| 2025 | USA and China [18] | <ul><li>Hexadecane</li></ul> |
| Water-in-oil emulsion | | |
| 2010-2018 | Japan [13,14,15] | <ul><li>Cyclohexane</li><li>Dodecane</li><li>Benzene</li></ul> |

| | | • Octane |
|---|---|---|
| | | • Hexane |
| | | • Oleic acid |

The first hypothesis for explaining this paradoxical longevity would be the presence of impurities in the oil phase. Such impurities could serve as a surfactant, reduce surface tension, and so promote stability. There have been publications along these lines for both nano-bubbles and pristine emulsions. However, detailed verification experiments conducted by many groups with very thorough purification confirmed that longevity remains independently of the degree of cleaning. We provide several additional arguments against this hypothesis in Appendix 1.

A second hypothesis is that an electric surface charge at the pristine interface contributes to the observed stability. There is seemingly consensus on this. However, there is still some uncertainty regarding the origin of the surface charge. Most of the authors point towards adsorption of $OH^-$ ions as the major factor with vast supporting evidence. Nevertheless, there are alternatives.

First of all, the role of ions produced due to absorption of $CO_2$ remains unclear. Pristine emulsions could be considered as aqueous carbonated systems when they are open to air [23]. In addition to standard water species $H_2O$, $H^+$ and $OH^-$, they contain dissolved $CO_2$, carbonic acid $H_2CO_3^-$, bicarbonate anion $HCO_3^-$ and carbonate anion $CO_3^{-2}$. The last two anions could contribute to formation of electric surface charge on the pristine droplets interfaces when they are formed at natural atmospheric conditions. There are some indications mentioned for instance in the paper [17] that this occurs indeed. Resolving this unambiguity is the first goal of this paper that we addressed experimentally.

We selected the same well-studied hexadecane-in-water pristine emulsion. In contrast to previous studies, we investigated evolution of this emulsion over time by measured pH, zeta potential and droplet size continuously over days and even weeks. We used acoustic spectroscopy for the droplet size measurement and electroacoustics for the zeta potential measurement; these methods allow us to study the emulsions without diluting them. These experiments yield serious support to the role of $CO_2$ related ions in charging the droplets.

Unfortunately, this is not sufficient for explaining pristine emulsions longevity. Existence of the surface charge and associated with it Electric Double Layer (EDL) cannot explain the mechanism of emulsification, which requires reduction of the surface tension. It is well-known that surface tension depends on ionic strength and existence of EDL according to the Onsager-Samaras theory [24,25]. However, the effect is very small, perhaps less than a few percent for pure water. It cannot explain the observed effect emulsification. There are important recent publications on relationship between water surface tension and electrochemical composition of water samples [26,27], but they do not amend this conclusion.

There is a hypothesis suggested for explaining existence of pristine emulsions by Eastoe and Ellis [28] – interfacial layer of the structured water molecules. They made reference to the remarkable 1945 paper by Frank and Evans [29]. These authors introduced an idea of water structuring when it gets into contact with non-polar substances. They wrote: "…*When a rare gas atom or a non-polar molecule dissolve in water at room temperature it modifies the water structure in the direction of greater crystallinity – the water, so to speak, builds a microscopic iceberg around it…*".

The same idea of structured water at interface was suggested by Bockris in 1963 [30], then experimentally verified with atomic force microscopy by Israelachvili and Pashley [31], theoretically supported by Derjaguin et al [32] and Mansui-Ruckenstein [33], and mentioned in the books by Lyklema [34] and Hunter [35]. Recent study of dynamic surface tension of water [26] and Raman spectroscopy of interfacial water layer [17] also indicates peculiarities in the water surface layer structure.

However, neither the structured water layer nor electric surface charge could not explain all the observed facts when taken separately. That is why we suggest extending the interfacial model created for nano-bubbles in the paper [22] to pristine emulsions.

Importantly, in addition to the structured water layer, this model includes the electric double layer caused by adsorbed anions. The combination of these two factors gives rise to two additional surface forces: dielectrostatic in the normal direction and repulsion of parallel water dipoles in the lateral direction. We extend this model further here and suggest applying it to explain the paradoxical existence of pristine emulsions.

**Materials.**

We conducted an experiment with the hexadecane-in-water emulsion similar to Beattie and Djerdjev [10]. Their method of developing pristine emulsion described in Ref [10] contains a lot of chemical purification procedures and prevention of $CO_2$ adsorption by using an $N_2$ atmosphere. The motivation of such an approach is clear – elimination of impurities as possible surfactants. However, it creates an impression that pristine emulsions are a rare and exotic fluke. The authors of that paper wrote that they were able to replicate their results under conventional conditions. That is what we wanted to test for removing a veil of mystery from pristine emulsions.

Therefore, we used initially hexadecane (HD) from Sigma-Aldrich with purity specified as ≥99%. We prepared several emulsions using this HD. We report the results of three of them, labeled as emulsions 1, 2 and 3. Then, after discussing results with several experts in the field, we were criticized that this HD is not sufficiently pure. Reference to the detailed Beattie and Djerdjev work [10] was not sufficient. Thus, we obtained cleaner HD on Sigma-Aldrich with purity specified as ≥99.8%. The potential amount of impurities is 5 times smaller. We calculated coverage of the droplet interfaces with this possible 0.2% impurities in Appendix 1. These calculations reveal that it is not sufficient for surface coverage. For water we used store-bought distilled water.

Water ionic strength was adjusted to 0.001 M using NaCl from Sigma-Aldrich.

The pH was adjusted using a 0.1 N solution of NaOH.

**Sample preparation and measurement protocol.**

We developed a different method of preparing emulsions because our goal was to study emulsion evolution, instead of only its electric surface properties as in the paper [10]. We also modified it along the way after learning more about the properties of these emulsions. Here is our preparation method and measurement protocol that we used for emulsion 4, with 99.8% purity HD.

There are two different experimental setups available for studying carbonated aqueous systems according to the handbook on aqueous chemistry [23]:

- "open", when $CO_2$ dissolves in water with no restrictions and its concentration in water is variable and ruled by Henry law [36]
- "closed", when the contact between water and air is blocked and the concentration of dissolved $CO_2$ is fixed.

We employ both experimental setups in this study.

***Open setup.***

In this section we describe procedure used for "open" setup when the sample preparation and measurements were conducted on open air, allowing full free equilibration of $CO_2$ content between air and the aqueous sample.

**Step 1.** Basic sample mixture. We added 3.7 g of hexadecane to 114.9 g of 0.001 M NaCl aqueous solution with high pH. This would ensure 4% vl emulsions assuming that we could mix these liquids. The initial pH value was close to 9 for emulsions 1, 2, and 3, and very close to 10 for emulsion 4. Emulsions 1,2 and 3 were prepared in separate glass vials. Emulsion 4 was prepared directly in the measurement cell of the measuring instrument.

**Step 2.** Formation of pristine emulsion. We sonicated the mixture for 1 minute using a high-power horn sonic probe, which allowed manual mixing of the sample and brought the top layers of the liquid into the sonic jet stream. This leads indeed to the formation of an opaque liquid with no visible phase border line.

**Step 3.** Measurement of mixture pH. This is a very important step that can be used for quick determination of whether the pristine emulsion could be formed. The pH value of the mixture must drop, reflecting adsorption of OH⁻ ions.

**Step 4.** Readjusting the pH. We returned the pH back to the initial value (9 - 10 range) by adding small amounts of NaOH. Emulsions must be mixed by magnetic mixer or pumping for ensuring the homogeneous spreading of NaOH.

**Step 5**. Homogenization. We sonicated for 1 more minute with the horn sonic probe.

**Step 6.** Equilibration for up to 20 hours. Our sample was left for equilibration while mixed with a low power agitator, a magnetic mixer. For emulsions 1, 2, and 3 we measured only pH during this equilibration because we did not know yet how important this short-term evolution is. In the case of emulsion 4 we measured pH, zeta potential and droplet size distribution continuously.

**Step 7.** pH readjustment. After the 20 hours equilibration we readjusted the pH back to the initial value (9 for emulsions 1, 2, 3 and 10 for emulsion 4). Then we sonicated the samples for 1 minute.

**Step 8.** Long term monitoring. We measured the droplet size distribution and zeta potential for 200 measurements, which took about 20 hours.

**Step 9 for emulsions 1, 2, 3**. Preservation for 5 days. We saved the emulsion after this set of measurements by pouring it back in the vial. The measurement of the same emulsion was repeated then in 5 days. During this period emulsions 1, 2 and 3 stay idle in the glass vial. They were mixed only during measurement when poured back into the measuring cell every 5 days.

**Step 9 for emulsion 4**. Preservation for 5 days. Emulsion 4 remained in the measuring cell and underwent continuous measurement of both the zeta potential and droplet size distribution. After the first day, continuous measurements were terminated because the zeta potential stabilized. But we kept measuring this emulsion for two more days, just 5 measurements per day for monitoring long term evolution of droplet size. Then we stopped because the droplet size reached a stable value.

### *Closed setup.*

The contact of the sample with outside air is terminated in this setup. As a result, the $CO_2$ content is supposed to be fixed during experiments. There is a simple test for verifying this condition – elevated value of pH up to 9 and higher must remain time independent even for pure water sample, with no emulsion droplets. We used this "elevated pH water test" for verifying possible adjustments for instruments cells and all our efforts failed. We could not close instruments cells to prevent air contact and instruments still remaining functioning.

Then we tried this test for water in various bottles with air tide cups. Surprisingly, initially we still observed pH drift from one day to the other. Then we realized this might be related to the $CO_2$ in the air above water level in the bottle if it is not completely filled. In order to verify this hypothesis, we filled 5 bottles with water with elevated pH to different levels, as shown on Figure

1. Bottle 1 is "fully filled", bottle 2 – ¾ filled, bottle 3 – ½ filled, bottle 4 – ¼ filled, bottle 5 – 1/10 filled.

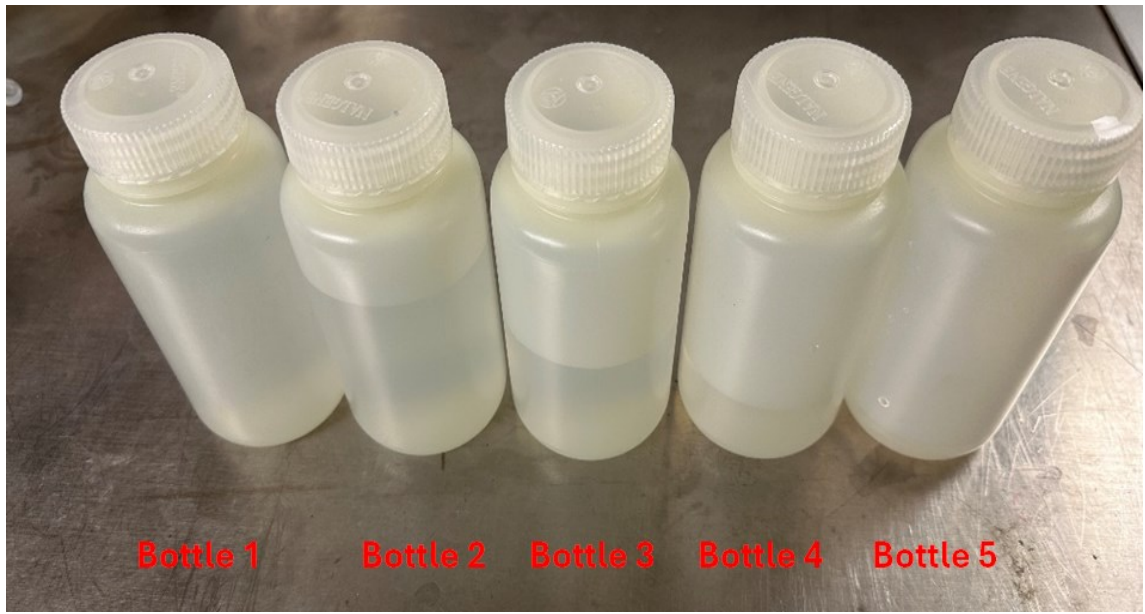

**Figure 1. Setup for verification "elevated pH water test" for fixed $CO_2$ content.**

We measured pH values in the bottles before closing them and then 24 hours later. We repeated this test twice, showing as run 1 and run 2 Figure 2 presenting measured pH values.

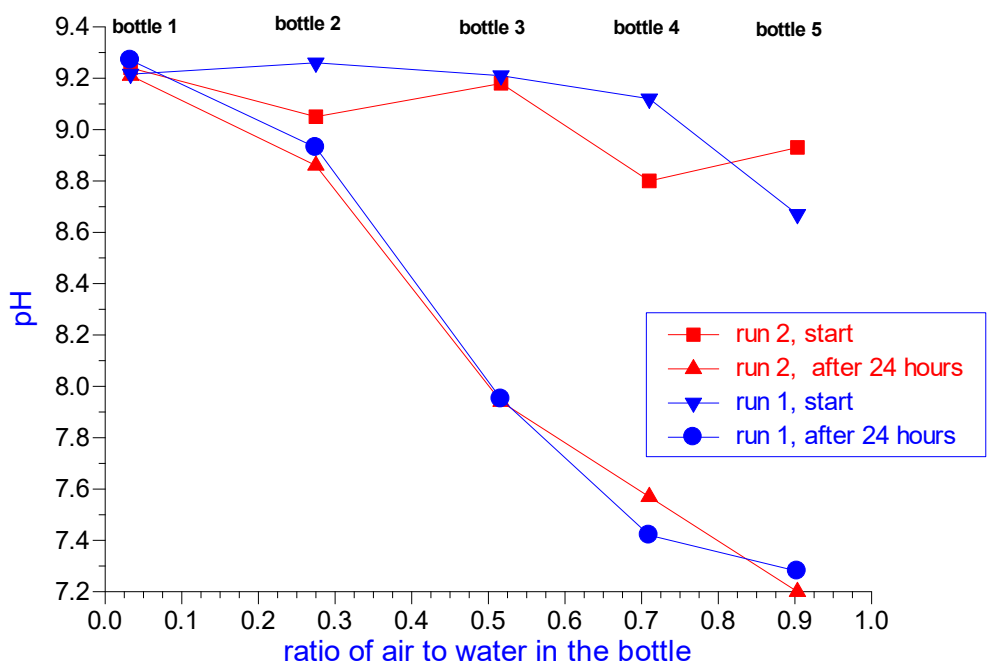

**Figure 2. Values of pH in the bottles before they are tidily closed and 24 hours later.**

The initial pH values differ because pH changes quickly for the open bottle with small water content. These variations are small. On other hand, variations of pH due to aging are much higher. These variations correlate exactly with ratio of water to air in the bottle: the more air and the less water leads to larger pH drop. This confirms hypothesis of $CO_2$ dissolving as the reason of such variations.

It is seen that pH remain time independent only for "fully filled" bottle. This leads to two necessary conditions for sample preparation for "closed setup":

- Sample cell must be tidily closed
- Sample cell must be "fully filled" with the sample.

These conditions limit significantly possible experimental setup. We were forced to prepare samples of emulsions in the bottles, same as in Figure 1. We did this by adding hexadecane to 0.001 M NaCl water solution with adjusted pH in the almost fully filled bottle. The amount of hexadecane was 4% vl of the total mixture.

Then we sonicated by inserting an ultrasonic horn into the bottle for 1 minute.

Then we measured pH and closed the bottle quickly.

For zeta potential measurements we took aliquot from the pipet from the middle of the bottle and filled cup on the face of the Zeta potential probe, as described in the book [34]. Then the cup for closed from the top for prevention $CO_2$ intake.

Measuring the droplet size distribution was conducted with filling sample fully into the measuring cell of the instrument.

**Instruments**

We used the Acoustic and Electroacoustic Spectrometer DT-1202. This instrument has an acoustic sensor for measuring ultrasound attenuation spectra within the frequency range from 3 MHz to 100 MHz. These spectra are the raw data used for calculating droplet size distribution. The details of our data analysis are given in Appendix 2.

This instrument also has an electroacoustic zeta potential probe that measures Colloid Vibration Current (CVI), which is the raw data for calculating droplet zeta potential. All details can be found in the book [37] and ISO standards [38-41]. This instrument can also measure conductivity, pH and temperature.

All specifications of these measurements, including accuracy and precision, can be found on the web site www.dispersion.com.

The functionality of pH probe was verified by measuring pH 10 buffer before and after every experiment.

The functionality of Zeta potential probe was verified by measuring zeta potential of silica Ludox certified reference material [42] before and after every experiment.

The functionality of the Acoustic sensor for droplet sizing was verified by measuring acoustic properties of water [37] before every experiment. Examples of droplet size distribution collected using this method are given in Appendix 2.

This instrument has built in magnetic mixer that creates extra pressure on the bottom of the cell, which in turn pumps liquid through the cell. It is sort of a centrifugal pump. It is possible to adjust the rate of pumping by setting magnetic cross rotation to different speeds.

We use ultrasonic probe by Sonic & Materials Inc operating at 20 KHz.

**Experimental data for "open setup".**

We discovered two different evolution patterns for this pristine emulsion for the "open setup". The short-term evolution occurs between sonication steps, and it shows in evolution of pH and zeta potential.

Figure 3 presents an example of pH short-term evolution. The value of pH was adjusted before each sonication step. Initial adjusted value was about 9.4 for emulsion 1,2,3 and close to 10 for emulsion 4. Following sonication, which triggers a change, the pH adjusts for about 20 hours.

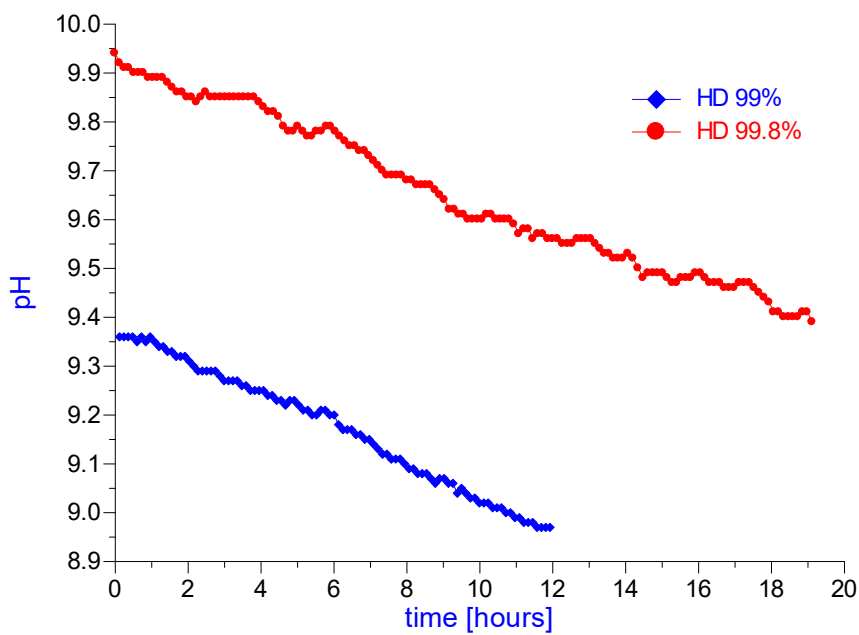

**Figure 3. Evolution of pH during the equilibration period. The blue line is for emulsion 3 with HD 99% purity, the red curve is for emulsion 4 with 99.8% purity. The initial adjusted value of pH (9.35 for emulsion 3 and 9.9 for emulsion 4) is different on purpose for determining the role of this adjustment level.**

The zeta potential variation correlates with pH variation during these short-term cycles between sonications, as shown on Figure 4. It is impossible to measure pH continuously during the second and third cycles because of the possible contamination. We have measured pH just before sonication and just before pH adjustment.

It is seen that the droplet diameter changes very little during the first cycle and is almost constant during the second cycle. Variation of the droplet diameter becomes more pronounced during the third cycle. This is the onset of the long-term evolution.

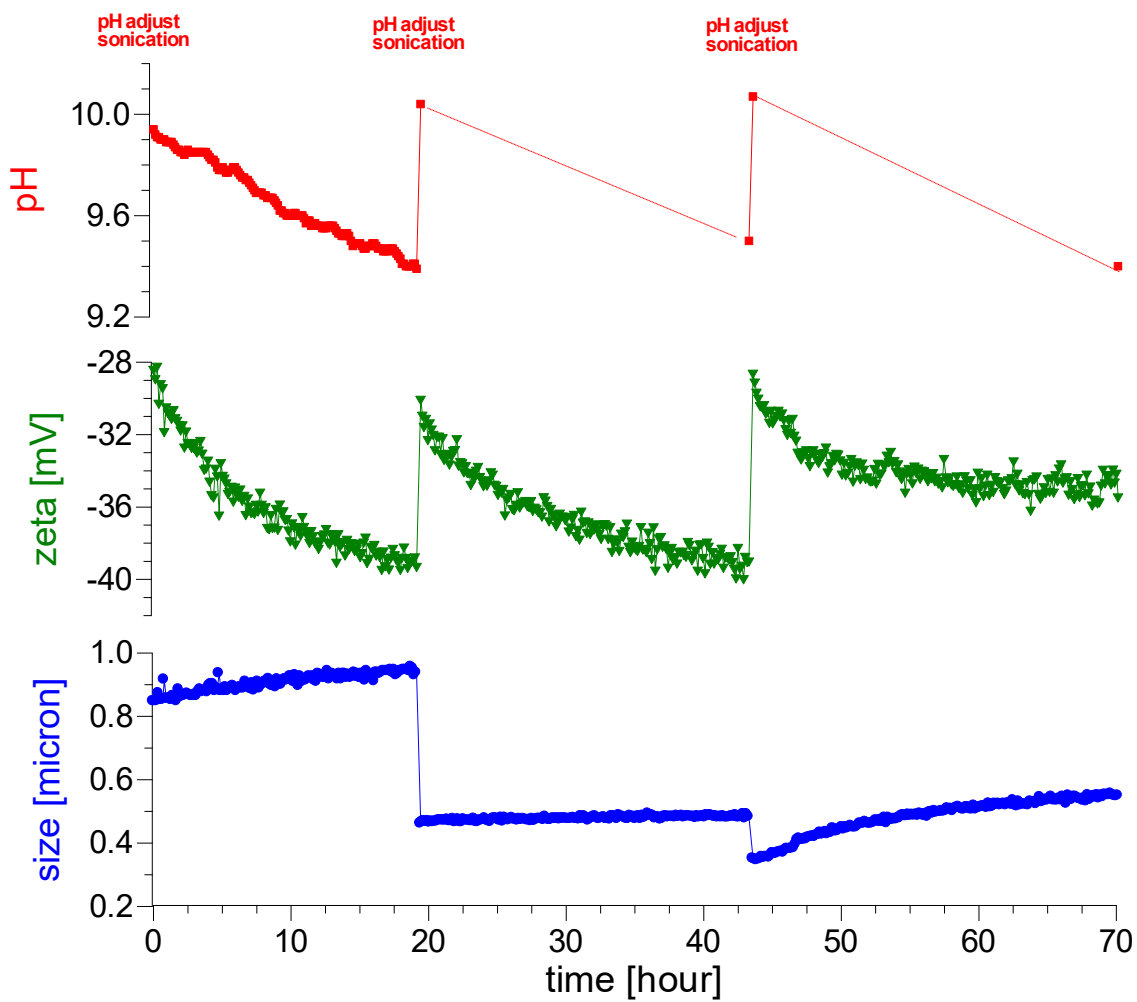

**Figure 4. Values of pH, zeta potential, and median dimeter of the emulsion 4 versus time, for the 3 steps of sonication with initial pH adjustment. Values of the pH could not be continuously measured during the second and third cycles because the measuring cell must be closed for preventing evaporation.**

The long-term evolution progresses after the first three days. It affects only droplet size. Zeta potential and pH remain constant. The evolution of the median droplet diameter is shown for all four emulsions in Figure 3. We have only a few points for emulsions 2 and 3, which are earlier tries, because we did not realize the existence of long-term evolution at that time.

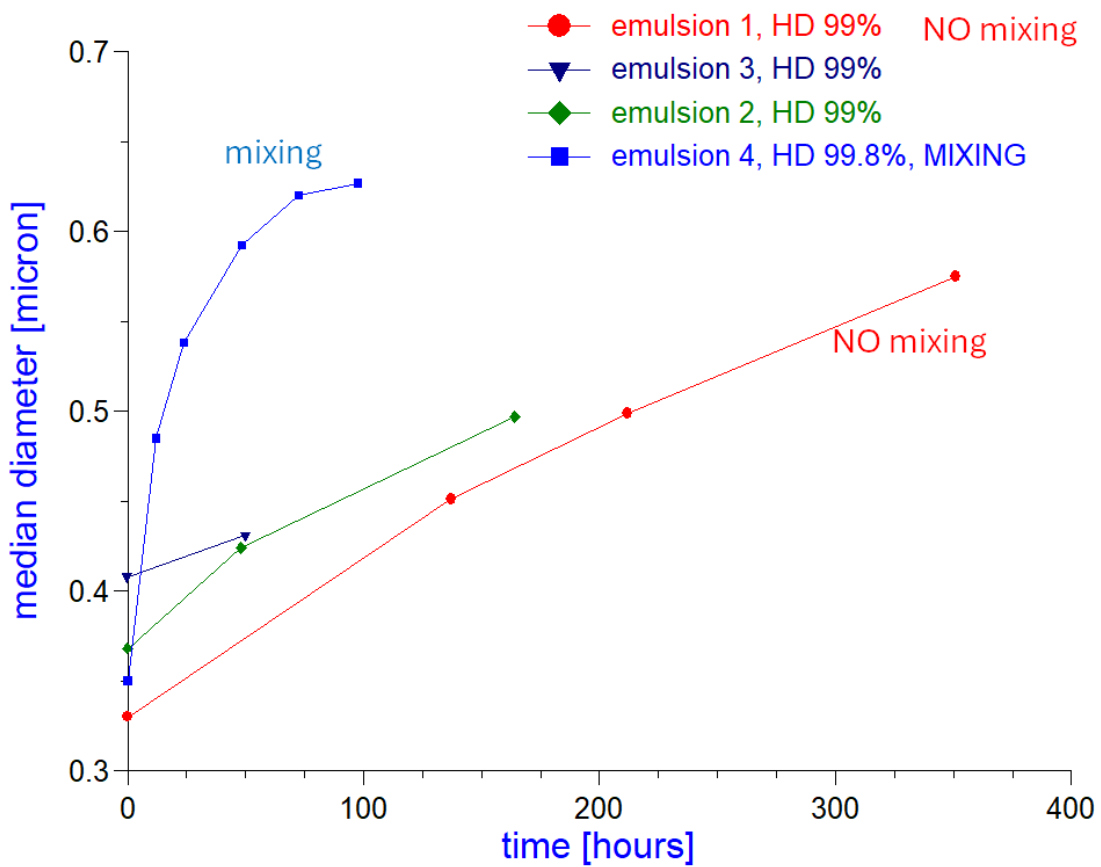

**Figure 5. Median diameter measured for all 4 emulsions over time. Emulsions 1 and 2 were measured before we recognized the long-term effect.**

It is seen that all these droplet size-time dependences merge towards the same value around 650 nm.

**Experimental data for "closed setup".**

In contrary to the "open setup" results, value of pH remains unchanged in the "closed setup" samples. We illustrate this statement with data in Table 2.

**Table 2. Values of pH for emulsion in "closed setup" over time.**

| Date | Value of pH for pH 10 buffer | Value of pH for emulsion |
|---|---|---|
| **Test 1** | | |
| Day 1 | 10.06 | 9.21 |
| Day 2 | 10.08 | 9.27 |
| Day 3 | 10.09 | 9.20 |
| **Test 2** | | |
| Day 1 | 10.08 | 9.66 |
| Day 2 | 10.04 | 9.53 |
| Day 3 | 10.11 | 9.55 |
| Additional sonication 1 min | | 9.49 |
| Day 4 | 10.07 | 9.46 |
| Additional sonication 1 min | | 9.51 |

There are other significant differences between both setups.

Measured values of electrophoretic mobility is 10 times smaller for "closed setup". This is shown in Figure 6. There are two reasons for explaining this decline of electrophoretic mobility. First of all, the volume fraction of droplets is much smaller because we could not emulsify the complete oil fraction into droplets. We do not know how much oil has been emulsified. That is why we used the volume fraction of the all added oil for calculating electrophoretic mobility, which makes it smaller. Secondly, the lower electric surface charge for "closed setup" would also lead to the decline of electrophoretic mobility. At this point we can only state that this result differs significantly from the electrophoretic data for the "open setup".

In addition, time dependence of electrophoretic mobility is quite different. It remains almost constant for the "open setup" at least on scale of tens of minutes. A single zeta potential measurement requires between 15 seconds for strongly charged particles up to 2 minutes for low charged particles. In contrary, electrophoretic mobility declines with time for the "closed setup". This decline is related most likely to the reduction of the droplets volume fraction near the face of the electroacoustic probe where electroacoustic signal is generated. The value of Colloid Vibration Current is proportional to the volume fraction of the dispersed phase [37]. If droplets are large then they move up due to buoyancy force, volume fraction drops, CVI declines as well, which show off as reducing electrophoretic mobility.

At the same time this decline confirms that reported electroacoustic signal is associated with droplets, and not with ions in the water phase.

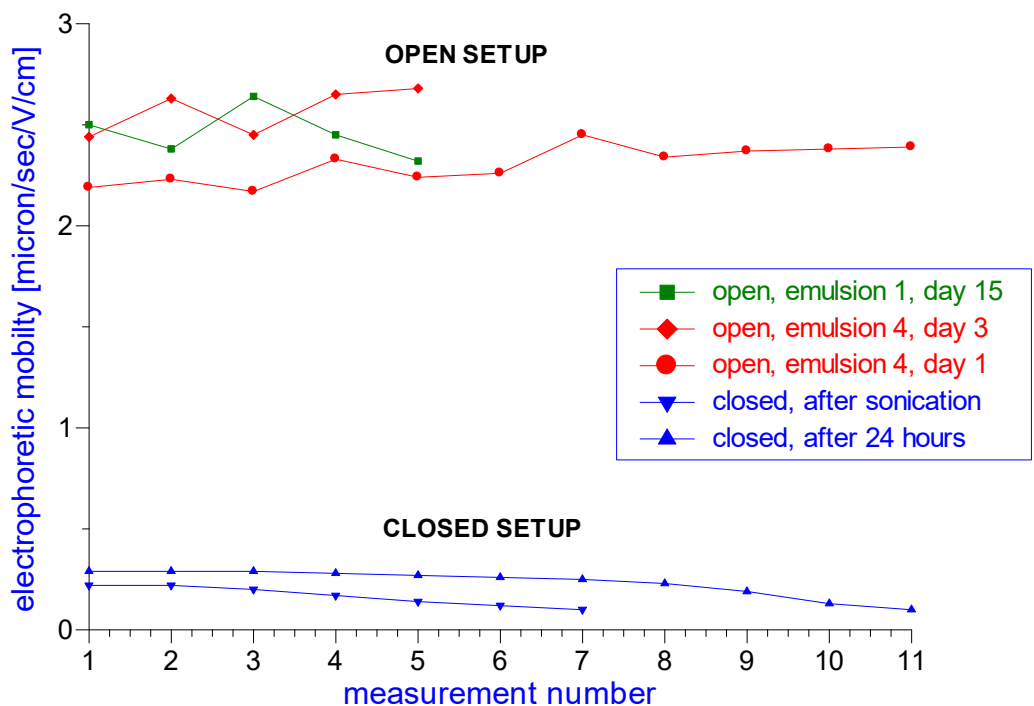

**Figure 6. Electrophoretic mobility of emulsions at different accesses to CO₂, in "open setup" and "closed setup".**

The droplet size measurement using acoustic spectroscopy was inconclusive for "closed setup". The raw data (attenuation frequency spectra) was very different comparing to the "open setup". It pointed out towards much larger droplets, but determining definite numerical values for droplets was impossible. Application of other methods for droplet size determination would be questionable because they require dilution, which would affect droplet size significantly.

**Discussion**

We observed two evolution patterns for hexadecane-water emulsion created by sonication for the sample that are exposed to air. We start by discussing the short-term process that evolves over tens of hours. This process is associated with declining pH and simultaneously increasing absolute value of zeta potential (Figure 4). This correlation indicates that some anions from the solution are being adsorbed gradually at the emulsion droplets interfaces. There are three anions in the original water solution that we used for making these pristine emulsions: $OH^-$ anions, bicarbonate anions $HCO^-$ and carbonate anions $CO_3^{-2}$. The last two exist due to dissociation of carbonic acid that occurs due to reaction of $H_2O$ with dissolved $CO_2$ from the air [23].

It would be logical to assume that absorption of $OH^-$ anions is responsible for pH decline, following the paper by Beattie and Djerdjev [10] and many other publications on pristine emulsions. They proved this by studying de-gassed emulsions with carbolic acid anions absent. There are, however, two arguments against this assumption.

**Argument 1.** It is not clear why bulk-interface equilibrium established over such long time – tens hours. There must be a factor that slows down adsorption of $OH^-$ ions but the nature of this factor is not clear.

**Argument 2.** Beattie and Djerdjev [10] observed zeta potential increasing with increasing pH. We observed the opposite trend – zeta potential increases with declining pH.

The other assumption about declining pH was suggested by reviewers of our initial paper. They pointed out that pH declined in the aqueous samples with elevated pH due to $CO_2$ dissolving in water if system is "open", accessible to air. This is a well-known fact in the chemistry of aqueous carbonated liquids [23].

This assumption resolves both Arguments formulated above.

Argument 1 about slow adsorption can be explained by the slow rate of $CO_2$ dissolving in the water.

Argument 2 about reverse Zeta potential pH dependence can be explained by adsorption of bicarbonate anions $H_2CO_3^-$ and carbonate anions $CO_3^{-2}$ at droplets interfaces. Increasing concentration of $CO_2$ over time leads to higher concentration of these anions and consequently higher adsorption which shows up as higher absolute value of zeta potential.

In order to collect more information on this emulsification process we conducted measurement at "closed mode", when content of $CO_2$ fixed and time independent. It turns out that we could not reproduce emulsion with properties same as for "open mode". Electrophoretic mobility is much lower; droplet size is much larger. The most importantly, value of pH remains constant.

First of all, this test confirmed that observed emulsification is not associated with impurities. Emulsification would be the same in both "open setup" and "closed setup" if impurities would be a leading factor.

Secondary, this test revealed that $CO_2$ and associated carbolic acid anions play an important role in charging the droplets. It seems that their adsorption contributes to the electric surface charge of hexadecane droplets. However, we cannot still determine which exact anion ($OH^-$, $H_2CO_3^-$, $CO_3^{-2}$) or perhaps some combination of them is responsible for the surface charging of the droplets.

The existence of the electric surface charge and corresponding Electric Double Layer (EDL) does not explain emulsions formation stability. It is well-known that surface tension depends on ionic strength, which had been explained by the Onsager-Samaras theory [24]. This theory takes into account long-range image forces of ion interactions with the interface and agrees with experimental data up to 0.1 M ionic strength. There is also a theory that fits experimental data for higher ionic strength, created by Ruckenstein and Mancui [25]. They considered an ion hydration effect and developed a theory for high ionic strength assuming the existence of ions free layer at the interface. The combination of the Onsager-Samaras and the Ruckenstein-Mancui theories agrees well with experimental data over the entire range of ionic strength. However, according to these theories the effect of electrolytes and the EDL on the surface tension is very small, perhaps less than a few percent for pure water. It cannot explain the observed effect of emulsification.

There is also a well-known Thomson theory for the role of electric charges in the droplet evaporation [43-45]. It shows that the presence of a charge on a droplet diminishes the evaporation tendency of the droplet because the electrostatic potential energy of the droplet increases as the droplet evaporates and more work has to be available to evaporate the charged droplet than when it is neutral. The Thomson effect is opposite to the well-known Kelvin effect associated with surface tension, which enhances the evaporation tendency of the droplet. Thomson derived an equation describing the equilibrium vapor pressure over a charged droplet of given radius, which is widely known as Kelvin-Thomson equation. However, it does not take into account screening of the surface charge with diffuse layer. The application of this theory is questionable for the large droplets with sizes that significantly exceed the Debye length. These droplets are electroneutral because a diffuse layer screens the electric surface charge. That is why it seems that the presence of the EDL is not sufficient to explain the observed existence of the stable droplet size.

The only explanation we can suggest is the existence of the structured water layer at interface. This is the same model discussed by Frank and Evans in 1945 [29] and suggested for pristine emulsions by Eastoe and Ellis [28]. In general, the idea of the structured layer at the interface is even older. There is a short overview of the papers on this subject in Ref [22].

However, neither the EDL only, nor structured water layer separately, can provide the required explanation. However, their interaction can do so. This was shown in Ref [22] for nano-bubbles. We invoke the same hypothesis here for deriving answers for the remaining questions about the observed pristine emulsion evolution. It is shown that interaction between the structured water layer and the electric double layer produces two additional forces acting on the interface. These additional forces could promote emulsification.

In addition, this hypothesis provides possible explanation of peculiarities of the "long-term evolution" that we observed on scale of days and even weeks. It is possible that the observed growth of the droplet size is associated with Ostwald ripening [46]. We conducted calculation of this effects rate in Appendix 3 and showed that it is very slow in the case of these emulsions and cannot explain experimental data.

In addition, the rate of the emulsion droplet size growth is linked to the mixing status of the emulsion, according to Figure 5. The characteristic time is much longer for emulsions 3 comparing to the emulsion 4. Emulsion 3 was idle for days between measurements. In contrast, emulsion 4 was continuously mixed with the magnetic mixer that is built into the measuring cell of the instrument. It seems that mixing accelerates the process that is responsible for this evolution.

Mixing the emulsion 4 with a magnetic mixer gives a weak effect. It cannot disturb the balance of surface forces. The only conceivable effect that it could cause is increasing the number of collisions between droplets. Therefore, we conclude that long-term evolution is due to coalescence. Droplets coalesce faster in the sample that is being mixed due to shear-induced collisions.

However, this coalescence is unusual. Normally coalescence leads to the gradual growth of the droplet size that ends with phase separation. Here we observe that the rate of coalescence declines with time, and it eventually stops leading to emulsion with stable diameter around 650 nm. Suggested hypothesis of the structured water layer in combination with EDL, as in the paper [22] predicts existence of such stable size. That is why we present here elements of this theoretical hypothesis applied to the studied emulsion.

There are two major elements of this hypothetical theory: "dielectrostatic force" and "repulsion of oriented water dipoles".

### *Dielectrostatic force determines stable size*

The first force is normal to the interface and competes with excessive pressure cause by interface curvature – Kelvin's effect in case of emulsions [43]. Its origin is the inhomogeneous EDL electric field acting on dipole moments of oriented water molecules. It was shown in Ref [22] that such force, called dielectrostatic, could compensate completely for the excessive pressure. The balance of these two forces leads to stable nano-bubbles size. We could assume that this balance also leads to the observed stable droplet size of our studied emulsions at the end of long-term evolution.

We can verify this assumption by calculating the radius of such stable droplets ($a_{stable}$) using the following Equation derived in Ref [22]:

$$a_{stable} = \frac{2\gamma}{\zeta * \kappa^2 * d_w * L_{str} * N_A * cw} \qquad (1)$$

where $\zeta$ is the zeta potential, $1/\kappa$ is the Debye length, $d_w$ is the dipole moment of a water molecule, $L_{str}$ is the thickness of the structured water layer, and $N_A$ is Avogadro's number.

We can apply this equation to predict a presumably stable droplet size of hexadecane-in-water emulsion that we study here. Here are the values of all involved parameters:

- $\gamma = 0.055$ N/m [47]
- $\kappa^2 \approx 0.003 *(1/9)*10^{20}$ 1/m² (for measured conductivity 0.032 S/m)
- $d_w = 6.17 * 10^{-30}$ C m
- $N_A = 6.02*10^{23}$ / mol
- $cw = 55,000$ mol/m³ for water

There are two parameters in Eq.1 with high degree of uncertainty – thickness of the structured water layer $L_{str}$ and electric potential at interface, assumed to be zeta potential $\zeta$.

We can assume that it consists of two water molecules layers and therefore $L=0.48 *10^{-9}$ m.

Electric potential at interface is actually much larger than measured zeta potential $\zeta$ because of the structured water layer, which is immobile. The measured absolute value of zeta potential is close to 40 mV. We assume here that the electric potential at interface is twice higher – 80 mV. This assumption would require verification in later studies, but it is sufficient now for approximate estimate.

Substituting these numbers into Eq.1 leads to the following result for diameter of the stable emulsion droplet $d_{stable}$ :

$$d_{stable} = 2 * a_{stable}$$

$$= \frac{2 * 2 * 0.055}{80 * 10^{-3} * 0.003 * \left(\frac{1}{9}\right) * 10^{20} * 6.17 * 10^{-30} * 0.48 * 10^{-9} * 6.02 * 10^{23} * 55500}$$

$$\approx 0.79 \; micron \qquad (2)$$

This is surprisingly reasonable value, close to the measured diameter of the stable droplet shown on Figure 3 as close to 0.65 micron. This is supportive of the hypothesis that the balance of normal forces between excessive Kelvin's pressure and dielectrostatic force might indeed control the droplet size.

There is one more aspect associated with the dielectrostatic force. Introduction of this force disrupts the balance between Kelvin's pressure and surface tension. There must be other factors that compete with surface tension at the stationary state. Such a factor was suggested in Ref [22] - repulsion of the water molecules dipole moments in the structured water interfacial layer. We suggest some calculation of this factor magnitude in the next section.

### *Repulsion of oriented water dipoles.*

The dipole moments of the parallel oriented water molecules in the structured surface layer repel each other contributing to the lateral interactions at interface. We can characterize it by contribution to the surface tension and assign symbol $\gamma_{str}$. We will try to estimate the value of this contribution and compare it with the known experimental value of water-hexadecane surface tension 0.055 N/m.

We begin with a general definition of the surface tension $\gamma$ that can be found in the Lyklema book [34], Eq. 2.4.5:

$$\gamma = \left(\frac{\partial U}{\partial A}\right)_V \qquad (3)$$

where U is energy, A is surface area.

Let us assume that one water molecule with dipole moment $d_w$ comes into this layer. It would cause change in the energy by $\Delta U_{dd}$ due to interacting with other molecules dipole moments and increase surface by $\Delta A$. Therefore, contribution to the surface tension is approximately equals:

$$\gamma_{str} = \frac{\Delta U_{dd}}{\Delta A} \quad (4)$$

In order to estimate the energy of the dipole-dipole interaction we consider interaction only with the molecular nearest neighbors of the added molecule. Figure 7 illustrates the simplest symmetrical positioning of water molecules in the element of the interfacial layer. We're using a simple square lattice and considering only the nearest 4 neighbors. Our calculation here is to check if the repulsion resulting from the parallel dipoles is at all comparable to the surface tension. The distance between molecules (L) is taken as constant. It is seen that the added molecule has roughly four neighboring molecules.

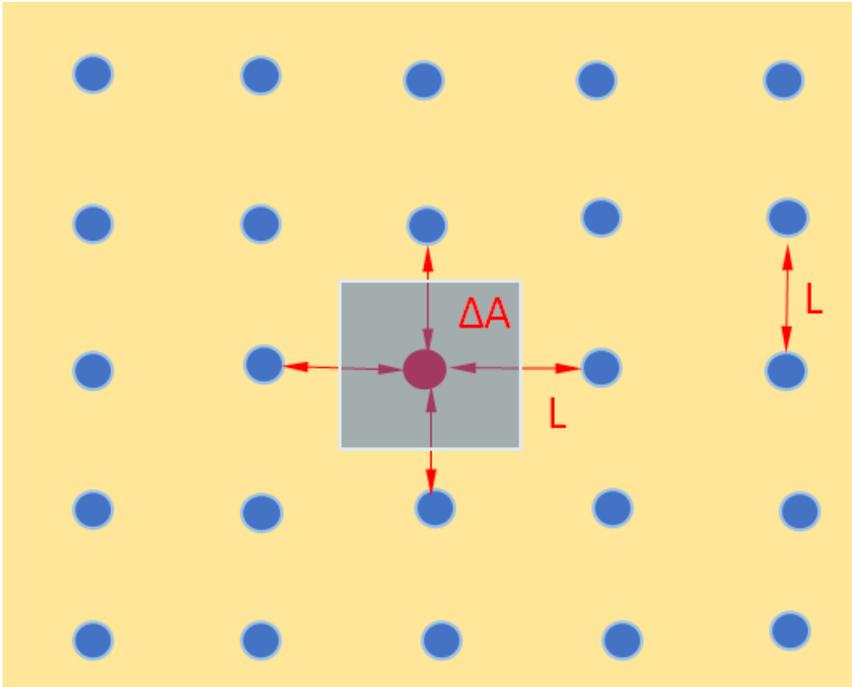

**Figure 7. Element of water-oil interface with blue circles symbolizing water molecules oriented perpendicularly to the interface. The red circle in the center is the added imaginary molecule. Shaded area around it illustrates ΔA.**

We also adopt simple additive approach to estimating interaction energy.

The interaction energy $U_{dd}$ of two parallel dipoles $d_w$ in a medium having dielectric constant ε is:

$$U_{dd} = \frac{d_w^2}{2\pi\varepsilon\varepsilon_0 r^3} \quad (5)$$

where $\varepsilon_0$ is dielectric permittivity of vacuum, r is distance between centers of the dipoles.

Assuming that the added molecule interacts with 4 others, the total change in surface energy due to this dipole-dipole interaction equals:

$$\Delta U_{dd} = 4 * \frac{d_w^2}{2\pi\varepsilon\varepsilon_0 L^3} \quad (6)$$

where we use r = L, according to Figure 5.

Variation in the surface area $\Delta A$ equals:

$$\Delta A = L^2 \quad (7)$$

Substituting Eqs. 6 and 7 into Eq. 4 leads to Eq 8:

$$\gamma_{str} = \frac{2 * d_w^2}{\pi\varepsilon\varepsilon_0 L^5} \quad (8)$$

The average distance between water molecules can be estimated from the fact that 1 $m^3$ of water contains 55,500 moles [(1000 kg / 0.018 kg/mol)]. The volume ($L^3$) corresponding to the single molecule equals 1 $m^3$ divided by $55500*N_A$. We find L from Eq 9:

$$L = \sqrt[3]{\frac{1}{55500 * 6.2 * 10^{23}}} \approx 0.3 \, nm \quad (9)$$

Now we can estimate the value of $\gamma_{str}$ :

$$\gamma_{str} = \frac{2 * d_w^2}{\pi\varepsilon\varepsilon_0 L^5} = \frac{2 * 6.17^2 * 10^{-60}}{3.14 * 80 * 8.85 * 10^{-12} * (0.3)^5 * 10^{-45}} = \frac{2 * 6.17 * 6.17}{3.14 * 8 * 8.83 * 3^5 * 0.1}$$

$$\approx 0.014 \, N/m \quad (10)$$

It turns out that potential contribution of oriented water dipoles repulsion to the surface tension can be close to the known value of hexadecane-water interface – 0.055 N/m.

This calculation underestimates dipole-dipole repulsion because we assumed the distance between them in the surface layer being the same as in the bulk – 0.3 nm. However, it should be smaller because surface tension pushing the closer. Reduction of this distance even by small amount would have a large impact on their repulsion due to $5^{th}$ power dependence. It turns out that there would be complete compensation of two effects if the average distance between water dipoles in the structured layer reduces to 0.23 nm, which seems quite reasonable.

### *Balance of forces at pristine interfaces with structured water layer.*

It is usually assumed that only two forces are acting at the water-oil-interface of an emulsion droplet: surface tension and Kelvin's pressure force due to curvature.

The existence of the electric double layer at the interface is an additional factor. However, it is not sufficient by itself to explain our experimental observation. That is why we employ the hypothesis of the structured water layer. This hypothesis leads to two more forces that counter-act with the classical ones. This new force balance is shown in Figure 8.

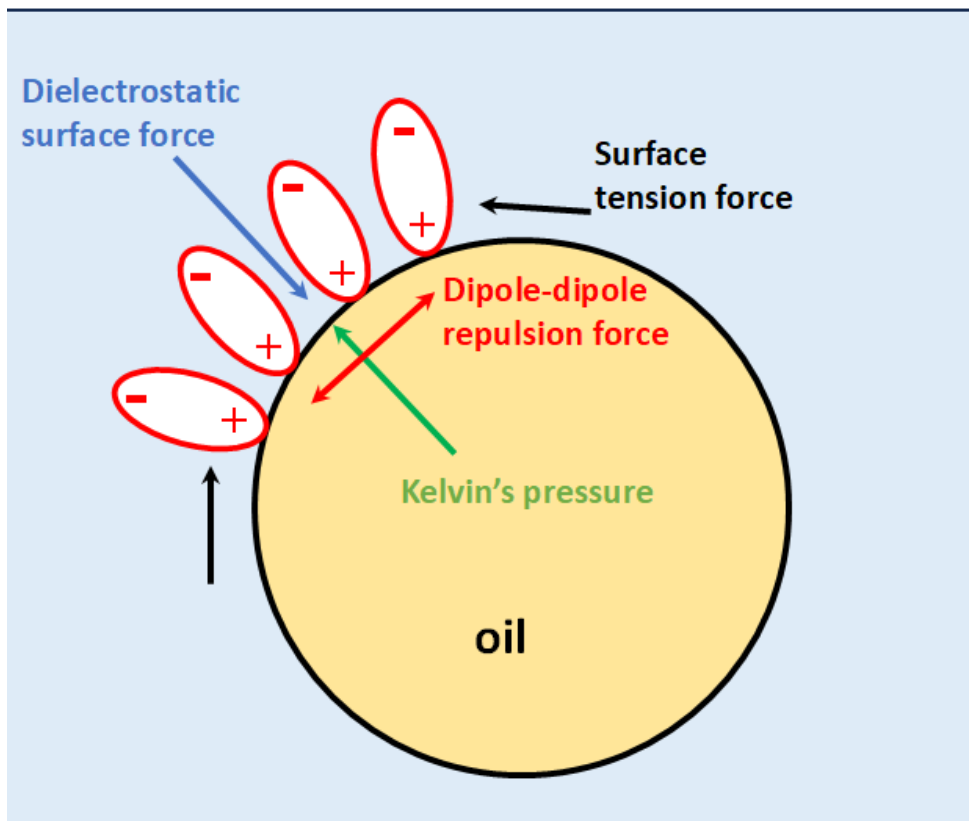

**Figure 8. Illustration of the force balance at pristine water-oil interface with structured layer of water molecules. Usually only two forces (Kelvin pressure and surface**

**tension) are considered; however, we are hypothesizing the existence of two additional forces not previously recognized (dielectrostatic force and dipole-dipole repulsion).**

There are two normal forces: Kelvin's pressure and dielectrostatic. They completely compensate each other at a particular droplet size, given with Eq. 2. There are two lateral forces at a given local position at the interface: surface tension and repulsion of water molecules dipole moments

### *What liquids could form "pristine emulsion".*

Not all liquids can form pristine emulsion. We tried, for instance, toluene and failed. It is possible that the capability of creating such pristine emulsion is linked simply to the small size of the oil phase molecule. Larger molecules can have a more stable emulsion. The molecular weight of toluene 92 g/mol is much smaller than molecular weight of hexadecane 226 g/mol.

There is an insightful analysis of the water molecule's structure building by non-polar molecules given by Frank and Evans [29]. They attributed insolubility of non-polar substances in water more to entropy effect, rather than internal energy. Here what they wrote: …"*Large, non-polar molecules have stronger van der Waals force fields around them than do small ones, and are more strongly held in any condense phase, including aqueous solutions. The larger they are, however, the larger the iceberg which they produce in water, and therefore the greater the loss in entropy involved in dissolving them…".*

It would benefit future studies of pristine emulsions to have a quick test that could predict whether a particular liquid is capable of creating such an emulsion. It seems that we have stumbled on such an experimental test when trying different methods of preparing hexadecane-in-water emulsion. Basically, it is a combination of Steps 1, 2 and 3 described above in the sample preparation procedure.

It seems that pristine emulsion would be possible if the pH of the mixture drops after the very first sonication. If it remains constant, above 9 as for the initial water used for the mixture preparation, then there is no adsorption of $OH^-$ ions, no interfacial charge change, and the mixture would not yield a stable emulsion.

It would be very interesting to conduct such quick tests with a wide variety of liquids and verify if the ability to form pristine emulsion correlates with molecular weight.

**Conclusions**

There have been multiple studies indicating that certain liquids form stable oil-in-water and water-in-oil emulsion with a longevity on the scale of days, even weeks without surfactants or any other surface stabilizing substance. We adopt the term "pristine emulsion" for these mixtures following Beattie and Djerdjev [10]. We reproduced emulsion from that study (4% vl hexadecane-in-water) by adjusting water pH above 9 and then applying sonication in several consecutive steps. We modified Beattie and Djerdjev approach somewhat. Instead of de-gassing it we did not isolate emulsion from $CO_2$. We used two different setups: "open setup" when sample is in non-restricted contact with air and $CO_2$ could dissolve in it freely, and "closed setup" when sample has fixed amount of $CO_2$ due to isolating it from air.

In the case of "open setup" we achieved full emulsification when all added hexadecane forms droplets. Such an emulsion remains relatively stable for hundreds of hours, but droplets properties exhibit certain evolution with two different patterns. The "short term" lasts about one day with pH and zeta potential changing and droplet size is practically constant. Interestingly, zeta potential absolute value increases with decreasing pH, which is exactly opposite to the results of the paper [10]. This is the first argument supporting the role of $CO_2$ in the observed emulsification.

The 'long term" pattern evolves much longer, for days and weeks with only droplet size changing and pH and zeta potential remaining constant.

In the case of "closed setup" when sample was isolated from air, we could not achieve complete emulsification, only fraction of added hexadecane formed drops. The value of pH remained constant and electrophoretic mobility was about 10 times lover than in the case of "open setup".

This difference in access to $CO_2$ for two different experimental setups indicates the importance of $CO_2$ in such an emulsification process. We cannot conclude yet which of the anions presented in the sample ($OH^-$, $H_2CO_3^-$ and $CO_3^{-2}$) adsorb at the droplets interface and form their electric surface charges. However, this surface charge and associated with it electric double layer play apparently a critical role in such emulsification.

The existence of electric surface charge does not explain the emulsification process because it does not ensure sufficient reduction of the surface tension. That is why we invoke hypothesis of the structured water layer at the interface, as was suggested for such emulsions previously by Eastoe and Ellis [28].

The structured water layer cannot explain the second evolution pattern, the long-term one, that we discovered. This long-term evolution is associated only with droplet size, and that growth occurs over several days or even weeks and then reaches a stable value. The zeta potential remains constant. The characteristic time of this pattern depends on the mixing conditions. That is why we attribute it to coalescence. However, this coalescence is unusual because it leads to a stable droplet size instead of phase separation.

Neither a "structured water layer" nor an "electric double layer" alone can explain such a peculiarity if considered separately. However, their interaction offers such possibility as was shown in Ref [22] for nano-bubbles, another object with a "pristine interface". It was shown there that the interaction of these two layers leads to two new surface forces – dielectrostic in normal direction and repulsion of water dipoles in the lateral direction. We suggest extending this model to pristine emulsions.

The dielectrostatic force could balance Kelvin's pressure force and determine observed droplet size of stable emulsion after long-term evolution. The calculated droplet size turns out to be close to the experimentally measured droplet size. The lateral force of repulsion between oriented water dipoles in the structed layer compensates for surface tension.

**Acknowledgment**

**Appendix 1. Arguments against hypothesis of impurities as origin of observed emulsification.**

First of all, several authors specifically mentioned that high reproducibility of their experiments ruled out impurities as potential explanation of emulsification.

Secondly, we also tested this hypothesis by multiple sonication steps. If surfactant-like impurities are responsible for the observed emulsion stability, then a simple one step sonication would produce such emulsions. Instead, we observed that multiple sonication steps with pH adjustments and equilibration are required. We describe the full procedure in the section on Materials and Sample preparation. A similar preparation procedure was described by Beattie and Djerdiev [10]. The impurities hypothesis cannot explain why these multiple steps are required for preparing pristine emulsions.

Third, we also discovered a large difference in emulsification depending on $CO_2$ content in water. The hypothesis of impurities cannot explain such differences.

Fourth, we design our experiments using hexadecane following the method of Beattie and Djerdjev [10], with >99.8% proof from Sigma-Aldrich. Here we present calculations indicating that 0.2% of surface-active impurity cannot provide sufficient surface coverage for emulsions with submicron droplet size. For all these reasons, we rule out the surfactant impurities hypothesis.

Let us calculate the amount of surfactant for complete coverage of the spherical droplet with radius $a_{drop}$. We assume that surfactant molecules can be characterized approximately with radius $a_{surf}$. Then the required volume fraction of surfactant $\varphi_{cr}$ equals to the ratio of the spherical layer around the droplet with thickness $2*a_{surf}$ to the droplet volume:

$$\varphi_{cr} = \frac{\left(4\pi d_{drop}^2 \, 2a_{surf}\right)}{\frac{4}{3}\pi a_{drop}^3} = \frac{6a_{surf}}{a_{drop}} \qquad (A2.1)$$

We have determined that the radius of the stable droplet size is about 300 nm, according to Figure 3.

For estimating the radius of the potential surface-active molecules that could be present in hexadecane samples as impurities we assume that it should have molecular weight similar to hexadecane. Otherwise, it would be eliminated by purification procedure. There is information of molecular size of substances with molecular weight that is close to the molecular weight of hexadecane (229) in the paper [48]. Radius of glucose molecules with molecular weight 180 is 0.33 nm, whereas radius of sucrose molecules with molecular weight 342 is 0.44 nm. Therefore, we can assume that the radius of the potential surface-active impurity $a_{surf}$ is about 0.4 nm. Substituting these numbers into Eq. A2.1 yields the following:

$$\varphi_{cr} = \frac{6 a_{surf}}{a_{drop}} = \frac{6 * 0.4}{300} = 0.008 = 0.8\% \qquad (A2.2)$$

The cleanest hexadecane that we use contains only potential 0.2% of impurities. It means that in the worst case when all impurities are surface-active, they could cover only 25% of the droplet interface. This number overestimates surface coverage because in reality, not all impurities are located at interface. A significant part would stay in the droplet bulk for maintaining surface-bulk equilibrium. Therefore, impurities could not stabilize interface and explain observed effects.

### Appendix 2. Details of particle sizing with acoustic spectroscopy.

The principles of Acoustic spectroscopy for particle sizing are described in the International Standards [38-41] and in the book [37].

Acoustic spectrometer measures ultrasound attenuation spectra in the range from 1 MHz to 100 MHz. Figure A.1 represents such attenuation spectra for 4%vl hexadecane-in-water emulsions measured in this study. We selected just 2 examples from hundreds of measurements conducted here. These two examples illustrate dependence of attenuation frequency spectra on the droplet size.

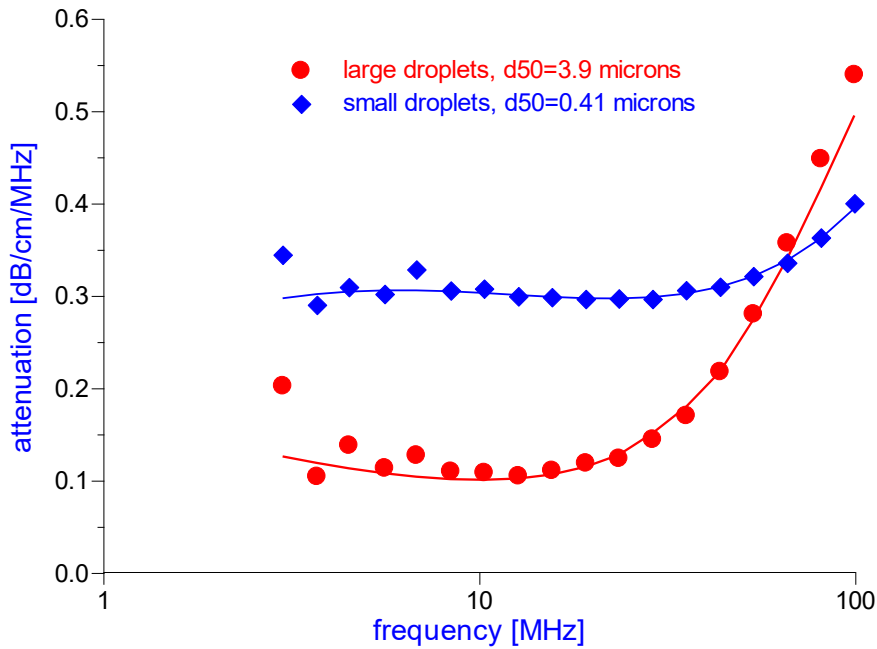

**Figure A.1. Attenuation frequency spectra for two 4% hexadecane-in-water emulsions at different stages of their preparation and consequently different droplet sizes. Symbols are experimental data points, curves are the best theoretical fit corresponding to the droplet size distributions shown on Figure A.2.**

Such attenuation frequency spectra serve as a raw data for calculating droplet size distributions. These calculations require, first of all, theoretical model describing attenuation of ultrasound in emulsions. There is a well verified experimentally version of the theory, see Chapters 4 and 6 in the book [37].

This theory provides a link between measured attenuation spectra and particle size distribution. It is seen that attenuation spectra are rather smooth functions of the frequency. Theoretical fitting of such smooth datasets does not allow extraction of large number of adjustable parameters without running into multiple solutions problem. In order to avoid this problem, we assume that droplet size distribution can be modeled as lognormal function. This reduces number of adjustable parameters for fitting attenuation spectra to just two: median size (d50) and standard deviation.

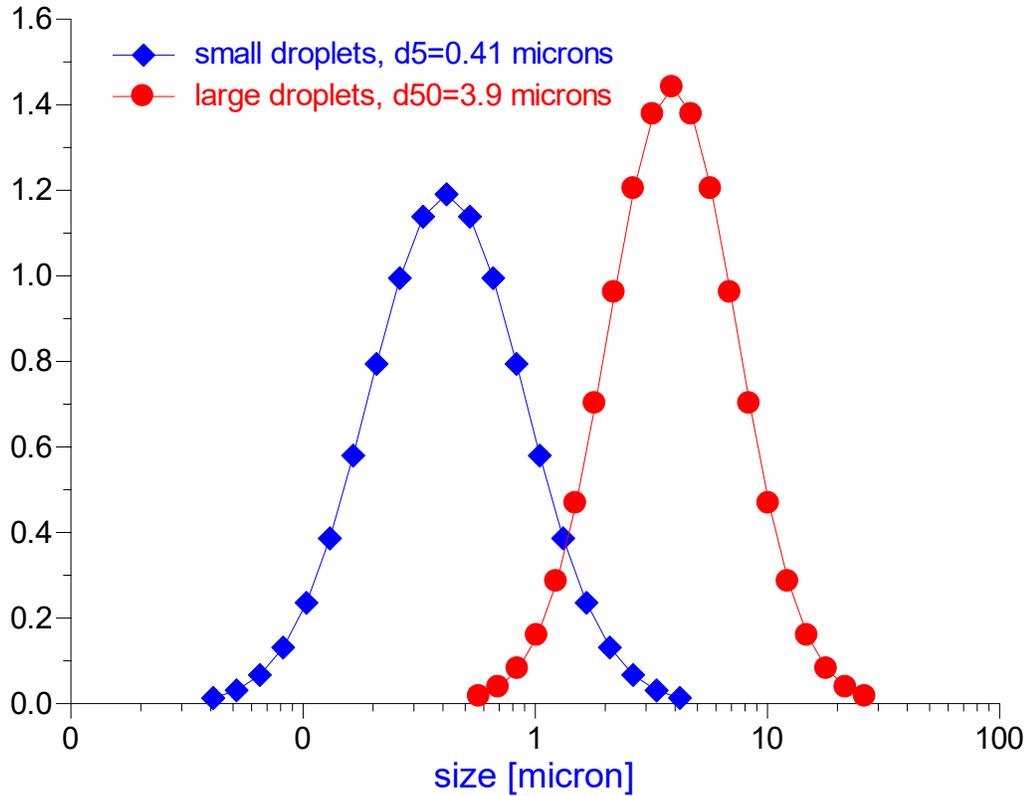

**Figure A.2. Droplet size distributions providing the best theoretical fit to experimental attenuation spectra shown on Figure A.1.**

Figure A.2 illustrates droplet size distributions that provide the best theoretical fit to experimental attenuation spectra, as shown on Figure A.1 with solid lines.

**Appendix 3. Calculation of Lifshitz-Slyozov-Wagner (LSW) [49,50] time for kinetics of Ostwald ripening [46].**

Ostwald ripening is a phenomenon observed in emulsions that involves the change of a droplet size over time due to transfer of emulsion droplet liquid from small droplets into the larger ones. This is a thermodynamically driven process caused by excessive pressure in the emulsion droplets due to curvature of their interfaces. It is usually referred to as Kelvin's pressure [43] and it is similar to Young-Laplace pressure in the case of bubbles [51,52]. This excessive pressure amplifies chemical potential of the molecules of the liquid that forms emulsion droplets. It becomes higher in the vicinity of smaller droplets compared to the vicinity of the larger droplets. Resulting diffusion flux moves these molecules from small droplets to larger ones. This description stresses the importance of polydispersity for Ostwald ripening [46]. However, it is possible to derive an equation for evolution of an average particle volume assuming some typical droplet size distribution. This was done by Lifshitz and Slyozov [49,50] who derived such equation for characterizing kinetics of Ostwald ripening:

$$\frac{da^3}{dt} = \frac{8DC_\infty \gamma M}{9\rho^2 RT} \qquad (A1.1)$$

Definitions of all symbols in this equation are:

- $a$ is an average radius of the droplets in [m]
- $t$ is time in [sec]
- $D$ is diffusion coefficient of the emulsion droplet liquid in [m$^2$/sec]
- $C_{sol}$ is solubility in [g/m$^3$]
- $\gamma$ is surface tension in [N/m]
- $M$ is molecular weight in [g/mol]
- $\rho$ is density in [g/m$^3$]
- $R$ is gas constant in [J/K$^0$ mol]
- $T$ is absolute temperature in K$^0$.

This theory yields following equation predicting value of the droplet radius at the time moment $t$ :

$$a(t) = a_0 \sqrt[3]{1 + t\frac{8DC_{sol}\gamma M}{9\rho^2 RTa_0^3}} \ = a_0 \sqrt[3]{1 + \frac{t}{t_{lsw}}} \qquad (A1.2)$$

where $a_0$ is an average droplet radius of the initial emulsion, at $t=0$.

Parameter $t_{lsw}$ equals:

$$t_{lsw} = \frac{9\rho^2 RTa_0^3}{8DC_{sol}\gamma M} \qquad (A1.3)$$

This parameter $t_{lsw}$ (LSW time) can be considered as characteristic time of Oswald ripening because an average droplet volume doubles during this time according to Eq.A1.2. Such definition of the characteristic time follows from the well-known theory of fast Brownian coagulation by Smolushowski [53]. Characteristic time is defined in that theory as the time required for double reduction in number of particles. In order to apply the same definition to Ostwald ripening theory we should convert it to particles volumes, because this is the main parameter changing over time due to the diffusion transfer of liquid from small droplets to the large ones. The reduction of particles number by factor of 2 can be achieved if particles volume increased by 2 times for the same total volume fraction. That is what we will use as characteristic time: ***time that is required for average particle volume to double, or an average droplet size increase by factor of $2^{1/3} \approx 1.26$.***

Parameter $t_{lsw}$ agrees with such definition of the critical coagulation time according to the Eq. A1.2.

We can calculate LSW time for 4%vl hexadecane-in-water emulsion. It is possible to find values of all required parameters of hexadecane in literature [47,48, 53-57]. We could also assume the value of initial droplet radius of about 0.2 micron, or $0.2 *10^{-6}$ meter. The result is 38 days. This is much longer than the duration of our experiments. Therefore, we can conclude that Ostwald ripening does not affect observed emulsion evolution.

### References

1. Kamogawa K, Matsumoto M, Kobayashi T, Sakai T, Sakai H, Abe M, Langmuir, 15, 1913-1919, 1999.

2. Sakai T, Kamogawa K, Harusawas F, Momozawa N, Sakai H, Abe M, Direct observation of flocculation/coalescence of metastable oil droplets in surfactant-free oil/water emulsion by freeze-fracture electron microscopy, Langmuir, 17 (2), 255-259, 2001.

3. Kamogawa K, Akatsuka H, Matsumoto M, Yokoyama S, Sakai T, Sakai H, Abe M, Surfactant-free o/w emulsion formation of oleic acid and its esters with ultrasonic dispersion, Colloids and Surfaces, A, 180, 41-53, 2001.

4. Sakai T, Kamogawa K, Nishiama K, Sakai H, Abe M, Molecular diffusion of oil/water emulsions in surfactant-free conditions, Langmuir, 18 (6), 1985-1990, 2002.

5. Sakai T, Takeda Y, Mafune F, Abe M, Kondow T. Monitoring growth of surfactant-free nanodroplets dispersed in water by single-droplet detection, J. Phys, Chem., 107 (13), 2921-2926, 2003.

6. Sakai T., Surfactant-free emulsions, Current Opinion Colloid Interface Science, 13(4), 228-235, 2008.

7. Pashley RM, Effect of degassing on the formation and stability of surfactant-free emulsions and fine Teflon dispersions, J.Phys. Chem., B 107, 1714-1720, 2003.

8. Maeda N, Rosenberg KJ, Israelachvili JN, Pashley RM. Further studies on the effect of degassing on the dispersion and stability of surfactant-free emulsions. Langmuir, 20, 3129-3137, 2004.

9. Burnett GR, Atkin R, Hicks S, Eastoe J. Surfactant-free "emulsions" generated by freeze-thaw. Langmuir 20, 5673-5678, 2004.

10. Beattie JK and Djerdjev AM, The Pristine Oil/Water Interface: Surfactant-Free Hydroxide-Charged Emulsions, Angew. Chem. Int. Ed. 43,3568–3571, 2004.

11. Djerdjev AM, and Beattie JK, Hydroxide and hydrophobic terabutylammonium ions at the hydrophobe-water interface, Molecules, MDPI, 30, 875, 2025

12. Hwanbo SAH, Lee SY, Kim BA, Moon CK, Preparation of surfactant-free nano oil particles in water using ultrasonic system and mechanism of emulsion stability, Nanomaterials, 12, 1547, 2022.

13. Horikoshi S., Akao Yu, Ogura T, Sakai H, Abe M, Serpone N. On the stability of surfactant-free water-in-oil emulsions and synthesis of hollow $SiO_2$ nanospheres, Colloids and Surfaces, A, 372, 55-60, 2010.

14. Sakai T, Seo K. Colloidal stability of emulsifier-free water-in-oil emulsions: effect of oil property, J. Japan Soc Colloid Mater, 87(11), 1-6, 2014.

15. Sakai T, Oishi T. Colloidal stabilization of surfactant-free emulsion by control of molecular diffusion among droplets, J. of the Taiwan Institute of Chemical Engineers, 92, 123-128, 2018.

16. Alizadeh A, Huang Y, Liu F, Daiguji H, Wang M. A streaming-potential-based microfluidic measurement of surface charge at immiscible liquid-liquid interface, International Journal of Mechanical Sciences, 247, 108200, 2023

17. Xiong H, Lee JK, Zare RN, Min W. Strong electric field observed at the interface of aqueous microdroplets. The Journal of Physical Chemistry Letters, 11, 7423-7428, 2020

18. Shi L, LaCour RA, Oian N, Heindel JP, Lang X, Zhao R, Head-Gordon T, Min W. Water structure and electric fields at the interface of oil droplets, Nature, 640, 87-93, 2025

19. Pickering, Spencer Umfreville (1907). "Emulsions". *Journal of the Chemical Society, Transactions*. **91**: 2001–2021.

20. LaCour RA, Heindel JP, Zhao R, G, Head-Gordon T. The role of interfaces and charge for chemical reactivity in microdroplets, Journal of American Chemical Society, 147, 6299-6317, 2025

21. Meegoda JN, Hewage SA, Batagoda JH. Stability of Nanobubbles, *Environ Eng Sci*, 35(11), 1216-1227, (2018).

22. Dukhin AS, Xu R., A new approach to explain nano-bubbles paradoxical longevity, Colloids and Surfaces, A, 700, 134805, 2024.

23. Snoeyink VL, Jenkins D, Water Chemistry, John Wiley & Sons, NY, 1980

24. Onsager L, Samaras NNT, The Surface Tension of Debye-Hückel Electrolytes, *J Chem Phys*, 2, 628, 1934.

25. Mancui M, Ruckenstein E, Specific ion effects via ion hydration: 1. Surface tension, *Adv Colloid Interface Science*, 105, 63-101, 2003.

26. Hauner IM, Deblais A, Beattie JK, Kellay H, Bonn D, The dynamic surface tension of water, J Phys Chem Lett, B, 1599-1603, 2017

27. Beattie JK, Djerdjev AM, Gray-Weale A, Kallay N, Lutzenkirchen J, Preocanin T, Selmani A, pH and the surface tension of water, JCIS, 422, 54-57, 2014

28. Eastoe J, Ellis C. De-gassed water and surfactant-free emulsion. History, controversy, and possible application. Adv in Colloid and Interface Sci., 134-135, 89-95, 2007.

29. Frank HS and Evans MW, Free volume and entropy in condensed systems. III. Entropy in binary liquid mixtures; Partial molal entropy in dilute solutions; structure and thermodynamics in aqueous electrolytes. Journal of Chemical Physics, 13, 11, 507-532, 1945.

30. Bockris JO'M, Devanathan MAV, Müller K, On the Structure of Charged Interfaces, *Proc R Soc Lon Ser-A*, 274 (1356), 55–79, 1963.

31. Pashley RM, Israelachvili JN, Molecular layering of water in thin films between mica surfaces and its relation to hydration forces, *J Colloid Interf Sci*, 101(2), 500-514, 1984.

32. Derjaguin BV, Dukhin SS, Yaroschuk AE, On the Role of the Electrostatic Factor in Stabilization of Dispersions Protected by Adsorption Layers of Polymers, *J Colloid Interface Sci*, 115(1), 234-239, 1987.

33. Mancui M, Ruckenstein E, The Polarization Model for Hydration/Double Layer Interactions: The Role of the Electrolyte Ions, *Adv Colloid Interface Science*, 112(1-3), 109-128, 2004.

34. Lyklema, J, Fundamentals of Interface and Colloid Science, Volumes 1, *Academic Press*, 1993.

35. Hunter RJ, Zeta Potential in Colloid Science, *Academic Press*, 1986.

36. Henry W, Experiments on the quantity of gases absorbed by water, at different temperatures, and under different pressure. *Phil Trans R Soc Lond* 1803:93, 29–43

37. Dukhin AS, Goetz JP: *Characterization of Liquids, Nano- and Microparticulates, and Porous Bodies using Ultrasound Ed 3*, London, Elsevier, 2017.

38. ISO 20998-1, Measurement and characterization of particles by acoustic methods. Part 1. Concepts and procedures in ultrasonic attenuation spectroscopy, ISO, Geneva, 2006.

39. ISO 20998-2, Measurement and characterization of particles by acoustic methods. Part 2 Guidelines for linear theory, ISO, Geneva, 2012.

40. ISO 13099-1, Colloidal systems - Methods for zeta-potential determination - Part 1: Electroacoustic and electrokinetic phenomena, ISO, Geneva, 2012.

41. ISO 13099-3, Colloidal systems - Methods for zeta potential determination - Part 3: Acoustic methods, ISO, Geneva, 2014.

42. Dukhin A.S. and Riesel A. "Overlapping double layers in electrokinetics of concentrated dispersion", J. of Colloid and interface Science, 609, 664-674 (2022)

43. Sir William Thomson, On the equilibrium of vapor at a curved surface of liquid, *Philosophical Magazine*, series 4, **42** (282) : 448-452, 1871

44. Thomson JJ, Applications of Dynamics to Physics and Chemistry, 1st ed., Cambridge University Press, London, 1888.

45. Thomson JJ, Conduction of Electricity through Gases, Cambridge University Press, London, 1906.

46. Ostwald W. (1897). "Studien über die Bildung und Umwandlung fester Körper" [Studies on the formation and transformation of solid bodies] (PDF). Zeitschrift für Physikalische Chemise. **22**: 289–330.

47. Goebel A, and Lunkenheimer K, Interfacial tension of the water/n-Alkane interface, Langmuir, 13, 369-372, 1997

48. Haynes, W.M. (ed.). CRC Handbook of Chemistry and Physics. 95th Edition. CRC Press LLC, Boca Raton: FL p. 3-294, 2014-2015.

49. Lifshitz IM, Slyozov VV. The Kinetics of Precipitation from Supersaturated Solid Solutions. Journal of Physics and Chemistry of Solids. **19** (1–2): 35–50, 1961

50. Wagner C. Theory of the aging of precipitates by dissolution-reprecipitation (Ostwald ripening). Zeitschrift für Elektrochemie. **65** (7): 581–591. 1961.

51. Young T., An Essay on the Cohesion of Fluids, *Philos T R Soc Lond*, 95, 65–87, 1805.

52. marquis de Laplace, P.S., Traité de Mécanique Céleste, (Paris, France: Courcier), *Supplément au dixième livre du Traité de Mécanique Céleste*, 4, 1–79, 1805.

53. Smoluchowski M, Z. Phys.Chem, 92:129, 1917.

54. Klochko L., Noel J., Sgreva NR., Leclerc S., Metivier C., Lacroix D., Isaev M. Thermophysical properties of n-hexadecane: Combined molecular dynamics and experimental investigation". International communications in heat and mass transfer, 137, 106234, 2022.

55. ECHA; Hexadecane (CAS 544-76-3) Registered Substances Dossier. European Chemical Agency, 2015.

56. Coates M et al; Environ Sci Technol 19: 628-32, 1985.

57. Boron WF, Boulpaer EL, Medical Physiology: A Cellular and Molecular Approach, Elsevier, 2005.